\documentclass[12pt,preprint]{aastex}

\usepackage{subfigure}

\slugcomment{accepted for publication in {\it ApJL}}

\shorttitle{Gaia parallax offset}
\shortauthors{Jao et al.}

\begin{document}

\title{Distance Dependent Offsets between Parallaxes for Nearby Stars
  and Gaia DR1 Parallaxes}

\author{Wei-Chun Jao}
\affil{Department of Physics and Astronomy, Georgia State University,
 Atlanta, GA 30302, USA}
\email{jao@astro.gsu.edu}

\author{Todd J. Henry}
\affil{RECONS Institute, Chambersburg, PA 17201, USA}
\email{thenry@astro.gsu.edu}

\author{Adric R. Riedel}
\affil{Astronomy Department, California Institute of Technology, Pasadena, CA 91125, USA}
\email{arr@astro.caltech.edu}

\author{Jennifer G. Winters}
\affil{Harvard-Smithsonian Center for Astrophysics, Cambridge, MA 02138, USA}
\email{jennifer.winters@cfa.harvard.edu}

\author{Kenneth J. Slatten}
\affil{RECONS Institute, Chambersburg, PA 17201, USA}
\email{kslatten@cpinternet.com}

\and

\author{Douglas R. Gies}
\affil{Department of Physics and Astronomy, Georgia State University,
 Atlanta, GA 30302, USA}
\email{gies@chara.gsu.edu}

%%%%%%%%%%%%%%%%%%%%%%%%%%%%%%%%%%%%%%%%%%%%%%%%%%%%%%%%%%%%%%%%%%%%
\begin{abstract}
%%%%%%%%%%%%%%%%%%%%%%%%%%%%%%%%%%%%%%%%%%%%%%%%%%%%%%%%%%%%%%%%%%%%

We use 612 single stars with previously published trigonometric
parallaxes placing them within 25 pc to evaluate parallaxes released
in {\it Gaia's} first data release (DR1).  We find that the {\it Gaia}
parallaxes are, on average, $0.24 \pm 0.02$ mas smaller than the
weighted mean trigonometric parallax values for these stars in the
solar neighborhood.  We also find that the offset changes with
distance out to 100 pc, in the sense that the closer the star, the
larger the offset.  We find no systematic trends in the parallax
offsets with stellar $V$ magnitude, $V-K$ color, or proper motion.  We
do find that the offset is roughly twice as large for stars south of
the ecliptic compared to those that are north.

\end{abstract}


%%%%%%%%%%%%%%%%%%%%%%%%%%%%%%%%%%%%%%%%%%%%%%%%%%%%%%%%%%%%%%%%%%%%
\section{Introduction}
%%%%%%%%%%%%%%%%%%%%%%%%%%%%%%%%%%%%%%%%%%%%%%%%%%%%%%%%%%%%%%%%%%%%

The first wave of results from the {\it Gaia} astrometry mission was
released on 2016 September 14, in what is known as the {\it Gaia}
First Data Release (\citealt{Brown2016}, hereafter DR1).  With this
release, astrometry has entered a new era, where a few lines of SQL
script yield millions of parallaxes, rather than tens of thousands
from {\it Hipparcos}, or thousands from over a century of painstaking
work from both the ground and space that produced parallaxes for
individually-targeted stars.  Because of the new rich dataset, the
{\it Gaia} results will undoubtedly change the way astronomers
investigate the stellar population of the Milky Way, and how they
create samples of stars for in-depth studies.  Because of the broad
utility of the {\it Gaia} results, it is important to check that the
new measurements are consistent with what came before.  In this paper,
we compare DR1 results to the fundamental sample of nearby stars,
which motivated the pursuit of trigonometric parallaxes in the first
place.

%%%%%%%%%%%%%%%%%%%%%%%%%%%%%%%%%%%%%%%%%%%%%%%%%%%%%%%%%%%%%%%%%%%%
\section{Matching Stars and Parallax Comparisons}
%%%%%%%%%%%%%%%%%%%%%%%%%%%%%%%%%%%%%%%%%%%%%%%%%%%%%%%%%%%%%%%%%%%%

To compare the new {\it Gaia} parallaxes to previously measured
values, we first used the available DR1 search tools to extract stars
with parallaxes placing them within 40 pc.  The SQL script used was
{\it SELECT * FROM gaiadr1.tgas\_source WHERE parallax $>=$ 25}.  This
query resulting 3402 stars within 25pc. We use this result to match
3806 systems within 25 pc collected by us in the past 10 years
\citep{Henry2015}. This list contains all stars with published,
accurate (uncertainty $\leq$ 10 mas), trigonometric parallaxes placing
them within 25 pc as of January 1, 2015. This 25 pc horizon was
previously defined by the Catalog of Nearby Stars \citep{NSC} and has
been widely used in nearby populations studies. Presumably single
stars with no known stellar, brown dwarf, or planetary companions in
this list were selected to reduce contamination of the dataset by
stars with parallaxes suffering from astrometric perturbations.  The
search in DR1 was extended to 40 pc for cross-matching to ensure that
stars beyond 25 pc in DR1 would be included.  Stars with {\it
  Hipparcos} identifiers were matched first.  If a star had no {\it
  Hipparcos} identifier, coordinate matching was used with a radius of
1\arcmin, after adjusting for proper motions between the DR1 epoch of
2015 and the epoch of 2000 of these nearby stars.  The resulting list
of 612\footnote{There are a total of 869 systems in this list with new
  DR1 parallaxes.} stars and both sets of parallaxes used for the
comparisons is given in Table~\ref{tbl:data}.

For the pre-{\it Gaia} measurements, if multiple trigonometric
measurements were published for a given star, a weighted mean parallax
value was calculated.  Because parallaxes come from various sources,
we make three comparisons here: (1) DR1 vs.~weighted mean
trigonometric parallaxes from all available sources, (2) DR1 vs.~{\it
  Hipparcos} parallaxes \citep{Leeuwen2007}, and (3) DR1 vs.~YPC
parallaxes from {\it The General Catalogue of Trigonometric
  Parallaxes} \citep{YPC}, sometimes called the Yale Parallax Catalog,
or simply YPC.  The first two columns in Table~\ref{tbl:data} are
coordinates in epoch and equinox 2000, followed by three columns of
identifiers.  Columns 6--8 list proper motions, position angles of the
proper motions, and their references.  The majority of proper motions
are from the Tycho 2 catalog \citep{Hog2000}.  A few are from {\it
  Hipparcos} or our SuperCOSMOS-RECONS survey \citep{Hambly2001}, the
latter of which are presented here for the first time.  Columns 9 and
10 are Johnson $V$ magnitudes and references. Note that some stars
without Johnson $V$ have values converted from Tycho $B_T$ and
$V_T$\footnote{Details of the conversions will be discussed in a
  future paper that outlines how we construct template stars to get
  these conversions; for converted $V$ values we list the reference as
  ``this work''.}.  The 2MASS $K_s$ (99.99 indicates no value in
2MASS) and {\it Gaia} $G$ magnitudes are listed in columns 11 and 12.
Columns 13--16 give weighted mean parallaxes, parallax errors, the
number of parallaxes in the weighted means, and references.  The DR1
parallaxes and errors are listed in the final two columns.

\begin{figure}
\centering \subfigure[]{ \includegraphics[scale=0.3,angle=-90]{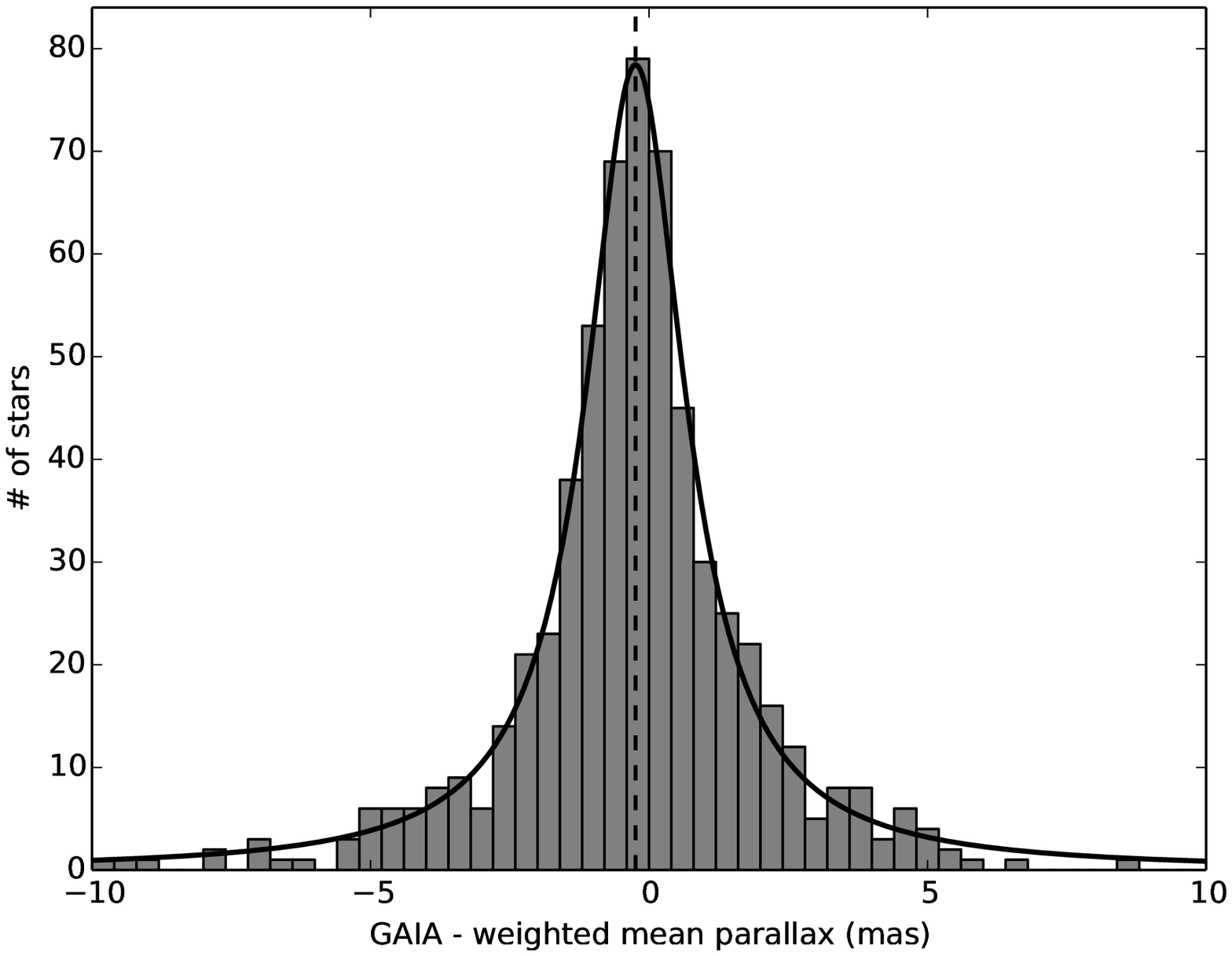}
\label{fig:histo}} 
\subfigure[]{\includegraphics[scale=0.3,angle=-90]{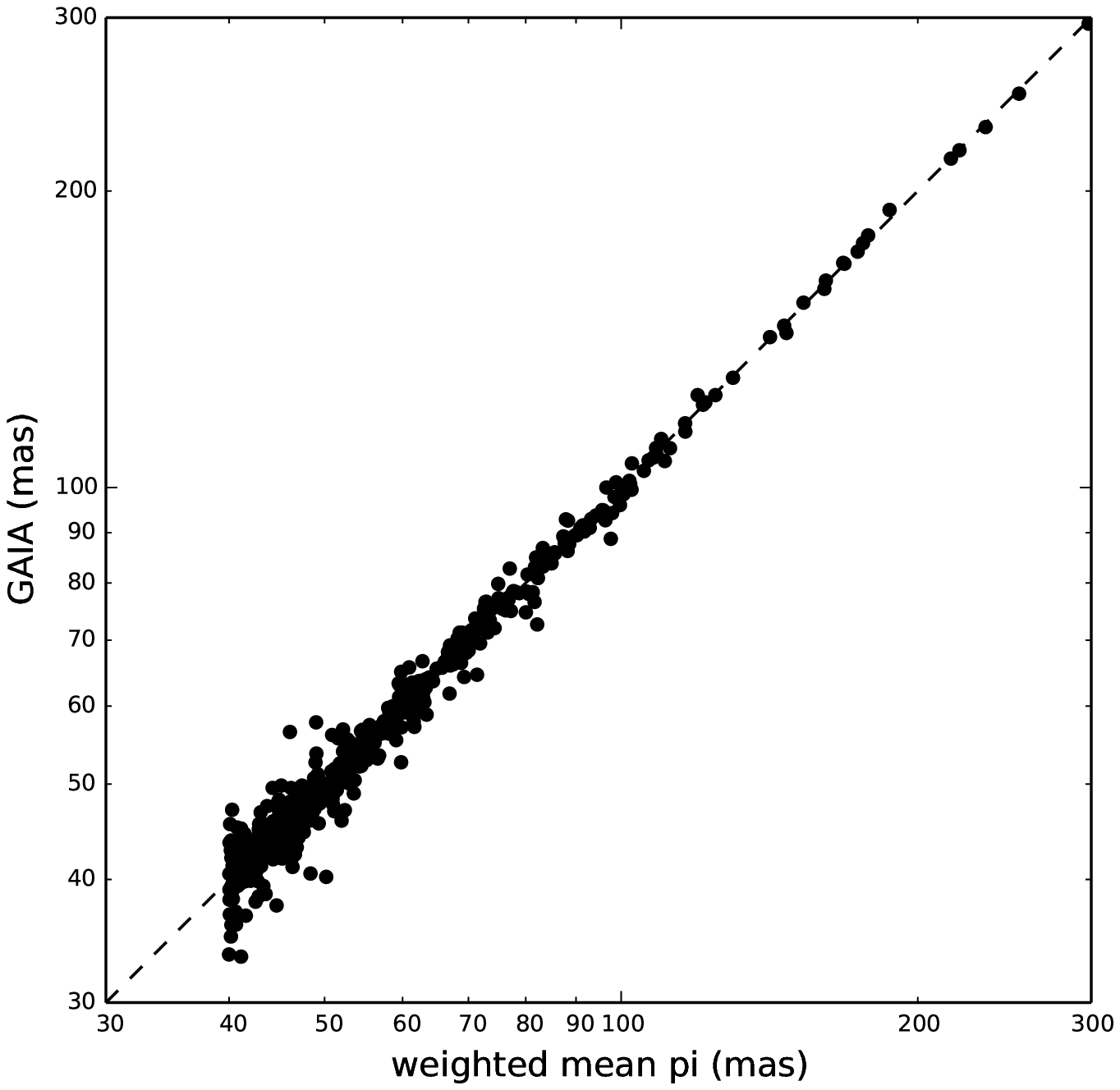}
\label{fig:one2one}} 
\subfigure[]{ \includegraphics[scale=0.3,angle=-90]{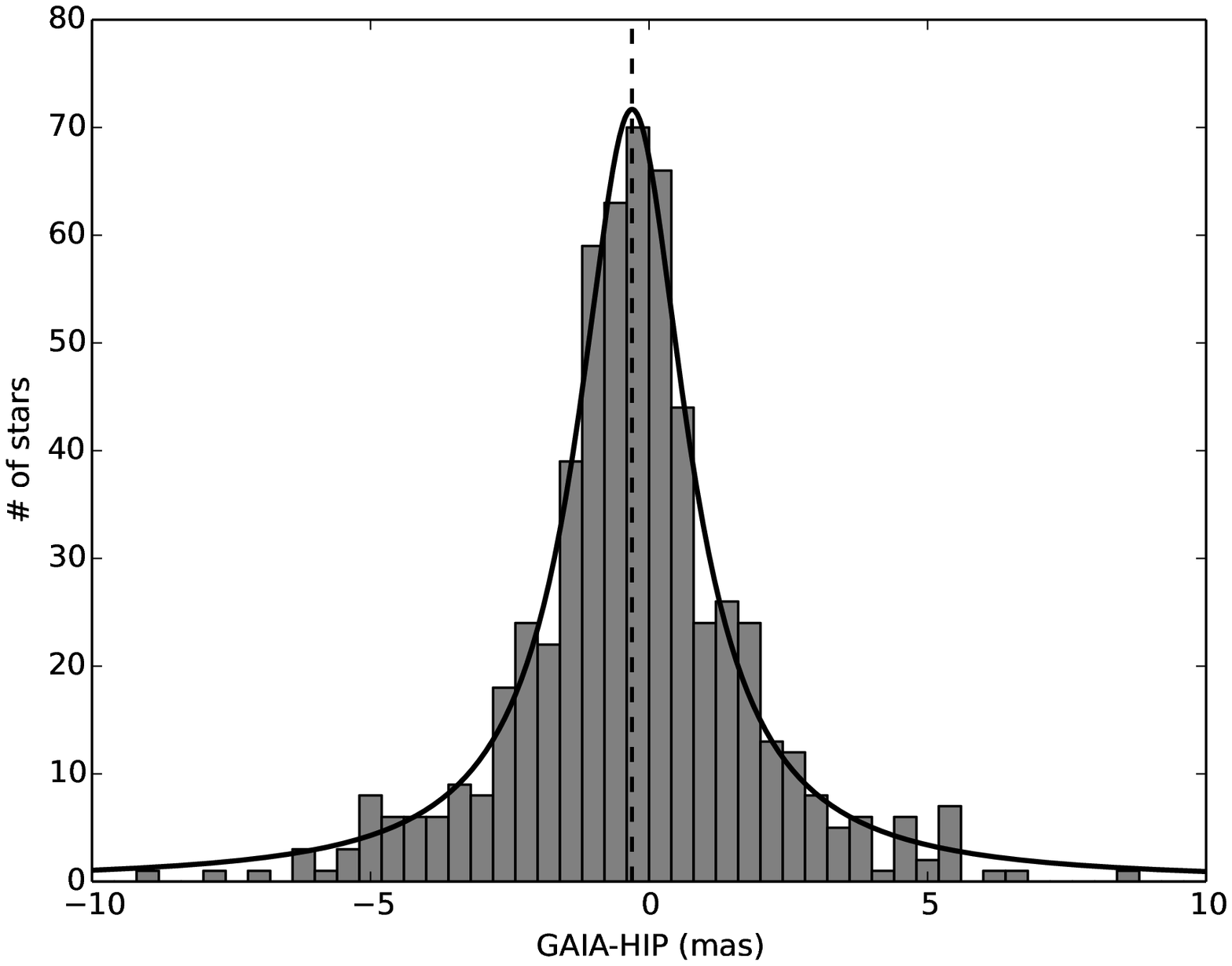}
\label{fig:HIPGAIA}} 
\subfigure[]{\includegraphics[scale=0.3,angle=-90]{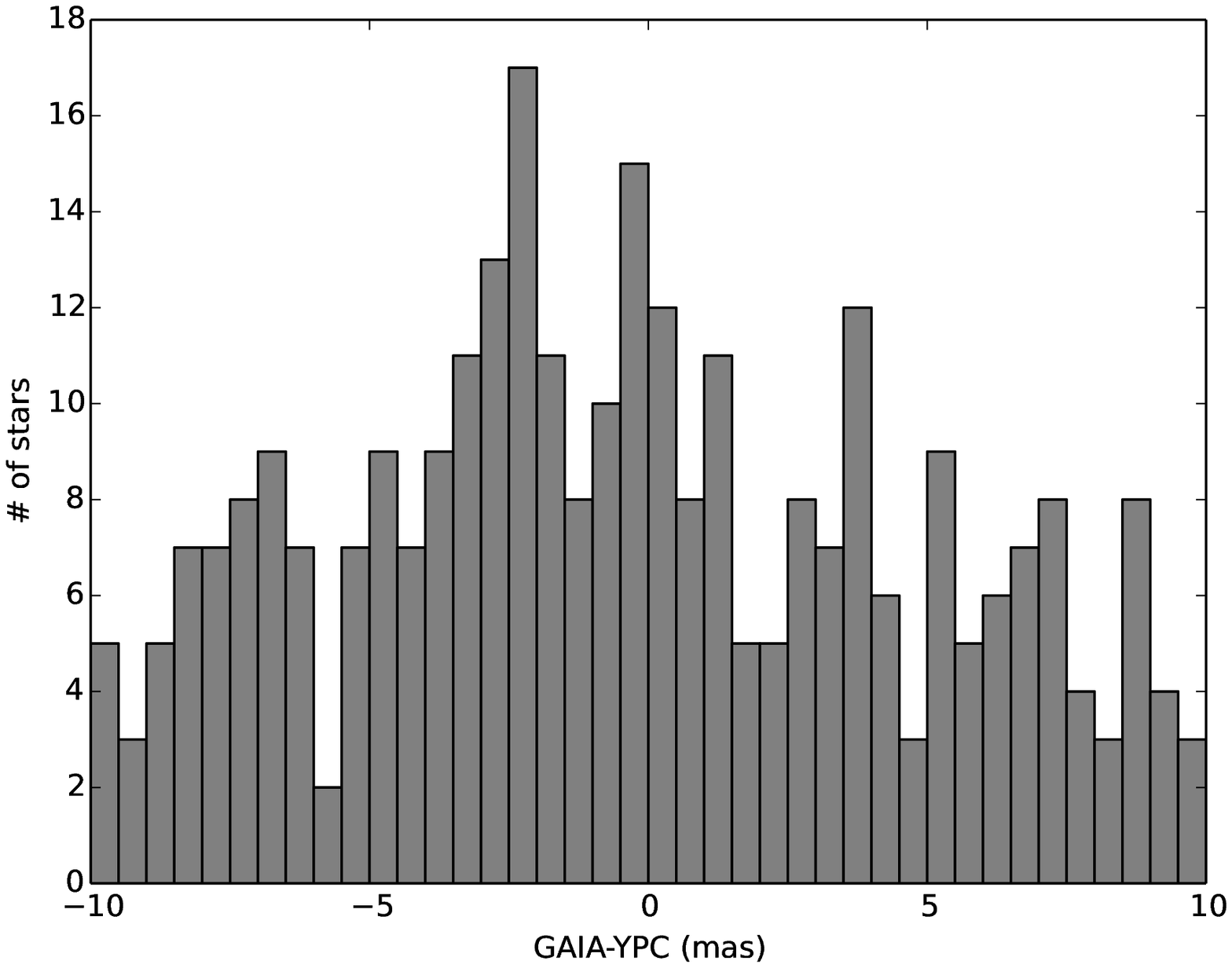}
\label{fig:YPCGAIA}} 

\caption{Parallax comparisons: \subref{fig:histo} A histogram of
  parallax differences, ($\Delta\varpi$), between values in {\it
    Gaia}-DR1 and weighted mean values for 612 single stars within
  25 pc is shown.  The bin size is 0.4 mas, and a Lorentz distribution
  is shown, rather than a Gaussian.  The dashed line indicates the
  offset of $-$0.24 mas between the two datasets. \subref{fig:one2one}
  This $\log-\log$ plot shows a one-to-one comparison between the DR1 and the
  weighted mean parallaxes.  The mean parallax errors for all points
  in DR1 and these 612 stars are $-$0.29 mas for DR1 and 2.33 mas for
  weighted mean parallaxes.  \subref{fig:HIPGAIA} A similar histogram
  to that shown in figure~\subref{fig:histo} is shown with a bin size
  of 0.4 mas, but for DR1 vs.~revised {\it Hipparcos} catalog parallax
  values.  The vertical line indicates an offset of $-$0.30
  mas. \subref{fig:YPCGAIA} A DR1 vs.~YPC comparison is shown with a
  bin size of 0.5 mas.  The distribution is much more stretched than
  for the weighted mean and {\it Hipparcos}-only distributions because
  of the larger YPC parallax errors.  Consequently, no fit has been
  made.  }
\label{fig:parallax}
\end{figure}

The Figure~\ref{fig:histo} shows binned results of the differences
between the DR1 and the weighted mean parallaxes, in the sense
$\Delta\varpi=\varpi_{DR1}-\varpi_{ weighted\ mean\ parallax}$.  The
Figure~\ref{fig:one2one} compares the parallax values along a
one-to-one correspondence line.  This histogram shows that the
$\Delta\varpi$ distribution is not symmetric about zero, but centered
around $-0.24\pm0.02$ mas, indicating that the DR1 parallaxes are
somewhat smaller than measured previously for nearby stars.  Because
of the unusually wide wings of the distribution, an unweighted Lorentz
distribution was fit instead of a Gaussian distribution that
represents random sampling.  The offset indicates that the DR1 results
place nearby stars systematically slightly further away than the
current measurements.  Our result is virtually identical to the
systematic offset of $-$0.25 mas reported in \cite{Stassun2016}, who
compared DR1 parallaxes to orbital parallaxes derived for 108
eclipsing binary systems.  In contrast, a comparison of DR1 parallaxes
to photometrically estimated parallaxes for 212 Galactic Cepheids
\citep{Casertano2016} does not show a systematic offset, instead
finding that the DR1 parallaxes are in ``remarkably'' good agreement
with the estimates from the period-luminosity relation
\citep{Riess2011}.  We note, however, that the result for Cepheids
involves photometric parallax {\it estimates}, rather than
trigonometric parallax {\it measurements}.

To probe possible sources of the offset, we also compared DR1 to {\it
  Hipparcos} and YPC results independently, as shown in
Figure~\ref{fig:HIPGAIA} and~\ref{fig:YPCGAIA}.  In
Figure~\ref{fig:HIPGAIA}, we see that the offset between DR1 and the
revised {\it Hipparcos} results \citep{Leeuwen2007} for 600 stars
common to both samples is $-0.30\pm0.03$ mas, similar to, and in the
same direction as the offset seen for the weighted mean parallaxes.
Again, a Lorentz profile has been fit to the distribution.  As shown
in Figure~\ref{fig:YPCGAIA}, the much larger errors in the YPC
parallaxes for 402 stars in common to both samples (offsets for 98
stars are beyond the edges of the panel) makes it difficult to reach
any clear interpretation about offsets between DR1 and YPC parallaxes.
Apparently, most of the differences seen between DR1 and weighted mean
parallaxes are due to offsets between {\it Gaia} and {\it Hipparcos}
results. 

\cite{Lindegren2016} also discussed parallax differences between {\it
  Hipparcos} values and the DR1 in their Appendix C.  They found that
the median parallax difference ($\varpi_{DR1}-\varpi_{Hipparcos}$) is
$-$0.089 mas based on 86,928 sources, roughly one-third of the offset
we have found.  Although they did not outline the distance
distribution of these 86,928 stars, analysis of their Figure B.1 for
$\sim$87000 stars shows that the majority have parallaxes between 0
and 10 mas.  Apparently, when the parallax comparison is done using
stars beyond 100 pc, the discrepancy is very small.

\begin{figure}
\centering 
\subfigure[]{ \includegraphics[scale=0.45,angle=-90]{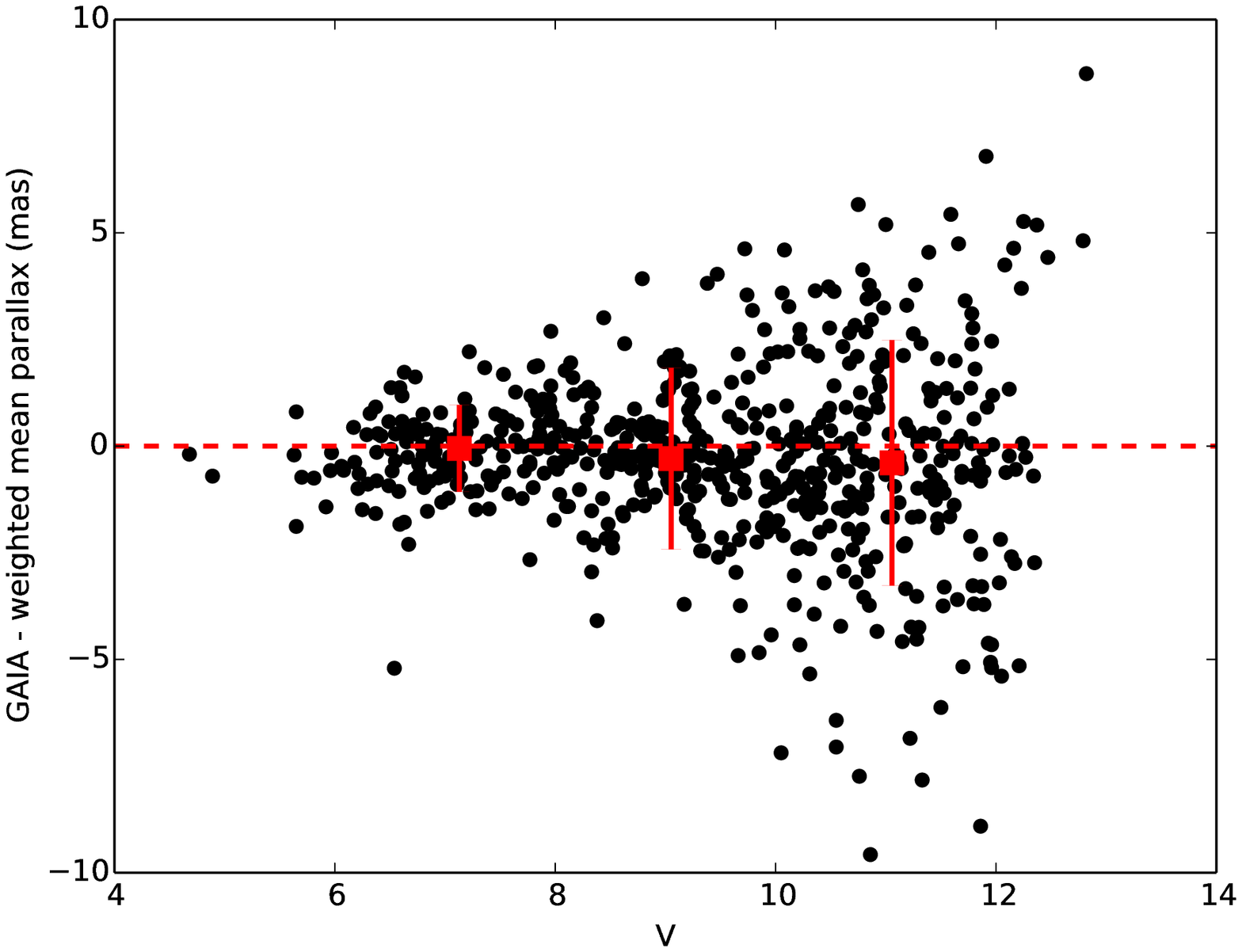}
\label{fig:magV}} 
\subfigure[]{\includegraphics[scale=0.45,angle=-90]{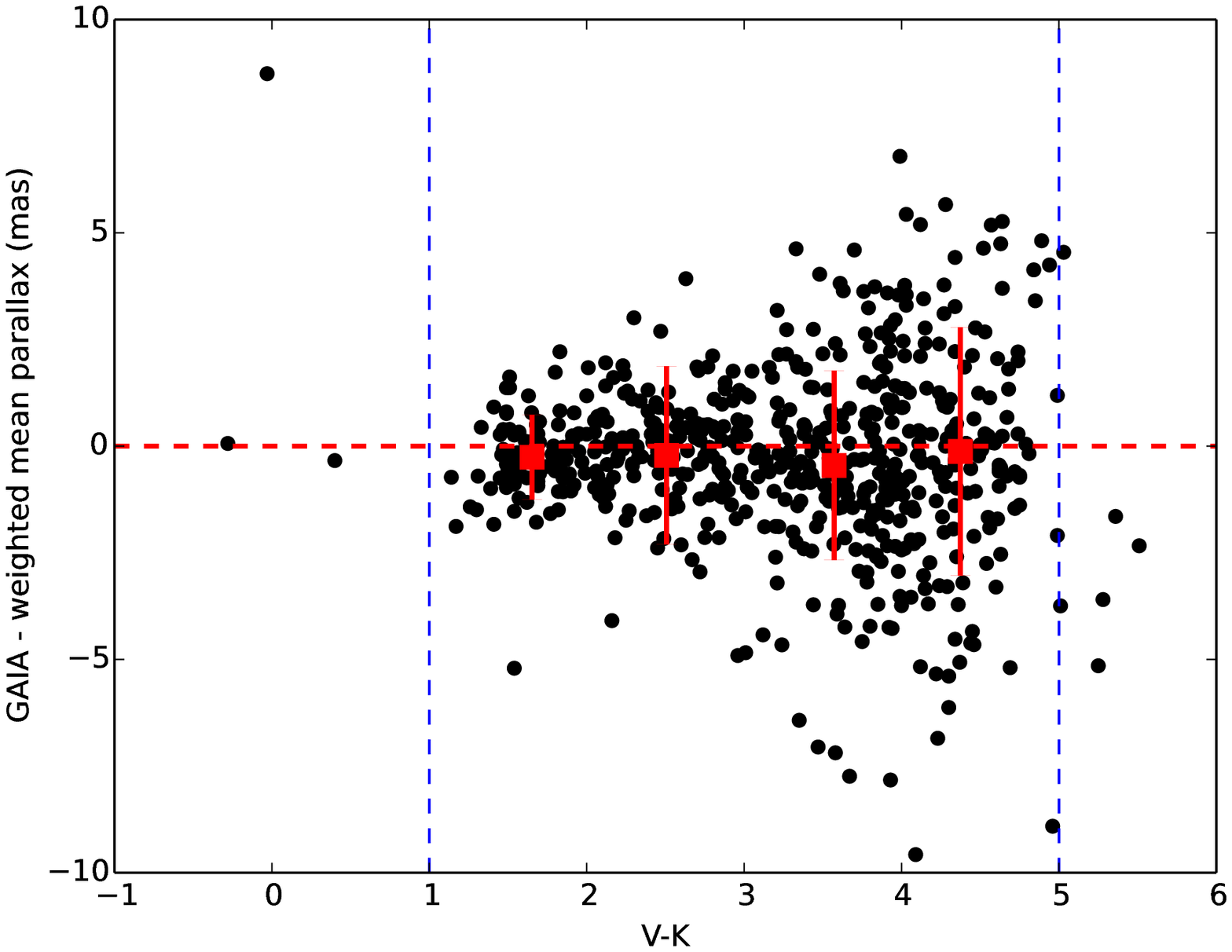}
\label{fig:color}} 
\subfigure[]{\includegraphics[scale=0.45,angle=-90]{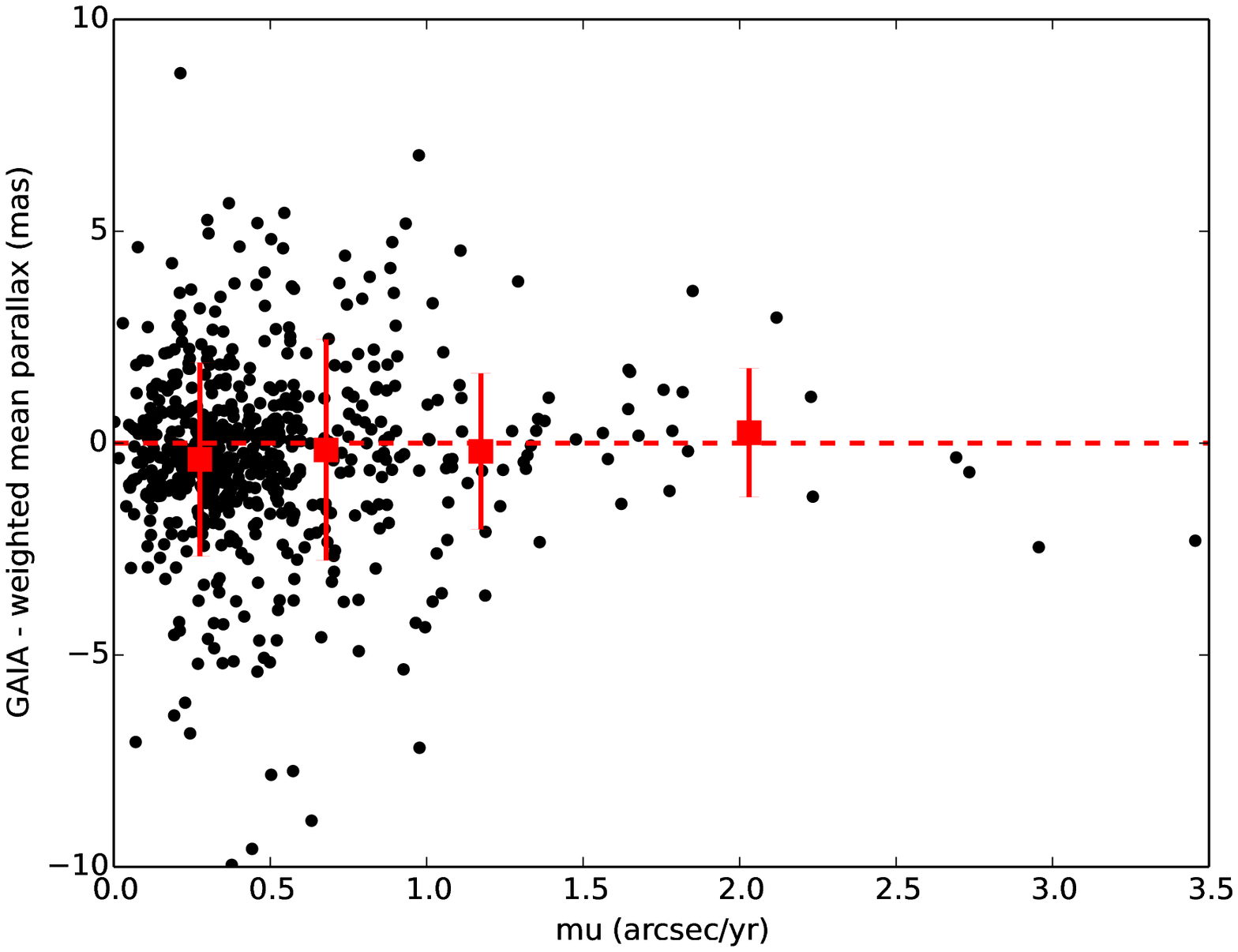}
\label{fig:mu}} 
\caption{The DR1$-$weighted mean parallax differences are plotted
  against $V$, $V-K$ and $\mu$.  The red dashed lines represent zero
  parallax differences.  The squares on each plot represent the binned
  valued for each of the parameters.  Because of the relatively large
  error bars, no obvious trends in parallax differences with any of
  these three attributes is evident.  Stars outside of two blue
  vertical dashed lines in the middle panel are not included when
  deriving the binned points in $V-K$.}
\label{fig:plot3}
\end{figure}

We also investigated the parallax differences in terms of $V$
magnitude, $(V-K)$ color, and proper motion to reveal whether or not
any of these attributes may be linked to the systematics.
Figure~\ref{fig:plot3} shows no clear trends between the DR1/weighted
mean parallax offset and any of these three parameters.  The only
evident correlations seen are expected --- stars fainter in $V$ and
redder in $V-K$ show larger offsets than brighter, bluer stars simply
because their previously measured parallaxes have larger errors in
general. In the DR1, the primary data set contains excess source
noises for each source, and the excess noises represent the modelling
errors \citep{Lindegren2016}. We found that the parallax difference is
not a function of the excess noise. We also note the Lutz-Kelker bias
has a negligible effect on this systematic offset reported here.

%%%%%%%%%%%%%%%%%%%%%%%%%%%%%%%%%%%%%%%%%%%%%%%%%%%%%%%%%%%%%%%%%%%%
\section{Systems Worthy of Note}
%%%%%%%%%%%%%%%%%%%%%%%%%%%%%%%%%%%%%%%%%%%%%%%%%%%%%%%%%%%%%%%%%%%%

Among the 612 stars used for our comparisons, there are two with
parallax differences larger than 20\%: GJ 723 and LTT 7370.

{\bf GJ 723} = HIP 91557 is a nearby M dwarf with parallaxes from YPC
and {\it Hipparcos}.  The weighted mean parallax is 46.12$\pm$5.81
mas, while the DR1 parallax is 56.483$\pm$0.267 mas.  The star has
highly discrepant parallaxes with large errors in YPC and {\it
  Hipparcos}.  The YPC parallax, 64.7$\pm$8.6 mas, is a weighted mean
of three independently measured parallaxes that are in reasonably good
agreement. The original {\it Hipparcos} parallax of 29.70$\pm$13.06
mas \citep{Perryman1997} was later revised to 30.49$\pm$7.89 mas in
the \cite{Leeuwen2007} re-reduction.  This shows the YPC parallax is
consistent with the DR1 parallax.

GJ 723 lies in a very crowded background star field, and the large
errors in the stochastic {\it Hipparcos} solutions may be due to close
proximity of GJ 723 to two other bright stars, 2MASS J18401792-1027481
and 2MASS J18401833-1027596.  GJ 723 has proper motions in R.A.~and
Decl.~of (-130.4, -523.0), (-170.2, -525.2), (-167.69, -520.08) and
(-143.6920, -557.0159) mas yr$^{-1}$, respectively, from the New
Luyten Two Tenths (NLTT) \citep{NLTT}, Tycho-2 \citep{Hog2000}, {\it
  Hipparcos} \citep{Leeuwen2007}, and DR1 catalogs.  The photometric
distance based on $VRIJHK$ colors \citep{Henry2004} puts the star at
21.46 pc, corresponding to a parallax of 46.59 mas that is larger than
the {\it Hipparcos} value and smaller than the YPC and DR1 values.
The distance to this star remains unclear.

{\bf LTT 7370} = HIP 91154 is a nearby K dwarf with a {\it Hipparcos}
parallax of 46.46$\pm$1.44 mas \citep{Leeuwen2007}.  No other
previously published parallax is available for this star, whereas the
new DR1 parallax is 26.083$\pm$0.762 mas.  The proper motions in
R.A.~and Decl.~from NLTT, Tycho-2, {\it Hipparcos}, and DR1 are
(-27.2, -389.0), (-18.6, -424.1), (-20.71, -424.84), and (-22.3672,
-424.9188) mas yr$^{-1}$, respectively.  The photometric distance
based on $VRIJHK$ colors \citep{Henry2004} places the star at 22.85
pc, corresponding to a parallax of 43.7 mas, which is consistent with
the {\it Hipparcos} value, but not DR1's. This field is not crowded.

When checking coordinates from {\it Hipparcos}, 2MASS, WISE, and DR1,
we found that these two stars move along their correct proper motion
tracks, as generally supported by the agreement of the proper motion
measurements given above.  Thus, there is no evidence that DR1 has
incorrect coordinates for these two stars.

%%%%%%%%%%%%%%%%%%%%%%%%%%%%%%%%%%%%%%%%%%%%%%%%%%%%%%%%%%%%%%%%%%%%
\section{Distance Dependent Offsets}
%%%%%%%%%%%%%%%%%%%%%%%%%%%%%%%%%%%%%%%%%%%%%%%%%%%%%%%%%%%%%%%%%%%%

The offset we describe above is a difference between DR1 and the
weighted mean parallaxes of these 612 presumably single stars within
25 pc.  We next investigated whether or not the offset depends on
distance, i.e.~the size of the measured parallax.  We separated the 25
pc sample into two sets --- stars with parallaxes between 40 and 80
mas and those with parallaxes larger than 80 mas --- and performed the
same histogram binning and Lorentz fitting routines for stars in these
two sets.  To reduce biases potentially caused by sample selection
near the 40 mas cutoff, i.e.~a subsample of stars beyond 25 pc in DR1
that were closer than 25 pc as measured previously, and vice versa, we
created the new subsets two ways: Method 1 used the weighted mean
parallaxes to define the parallax cutoff (612 stars) and Method 2 used
DR1 (580 stars).  Results for both methods are given in
Table~\ref{tbl:offset}, and show offsets of roughly the same size and
in the same direction.  Note that Method 1 was used to derive the
$-0.24\pm0.02$ mas offset value described above.

We then stretched the sample out to 100 pc to provide an extended view
of possible systematic offsets.  We used the following SQL script to
extract 16,260 parallaxes for stars within 100 pc in both
\cite{Leeuwen2007} and DR1: ``{\it SELECT parallax, parallax\_error,
  pmra, pmdec, plx, e\_plx, pm\_ra, pm\_de from gaiadr1.tgas\_source,
  public.hipparcos\_newreduction WHERE gaiadr1.tgas\_source.hip =
  public.hipparcos\_newreduction.hip AND parallax $>$ 10.}''  This is
considered Method 3.  Note that if stars are within 100 pc in
\cite{Leeuwen2007} but not in DR1, they would not be extracted using
this script.  We then used parallaxes from DR1 to separate stars into
different subsets corresponding to distance regimes of 25--50 pc and
50--100 pc, as presented in Table~\ref{tbl:offset}.  There is a clear
trend of changing offset in parallax with distance.  It is important
to note that we do not know the multiplicity status of the 16,260
stars used in Method 3.  Hence, many unresolved companions may cause
small perturbations and skew some {\it Gaia} measurements in DR1.  It
may be that the offset is much larger in Method 3 because unsolved
perturbations cause larger offsets in the DR1 parallaxes.  This table
also implies why the offset reported in \cite{Lindegren2016} is only
$-0.089$ mas because majority of those $\approx$87,000 stars are
beyond 100 pc.

Although \cite{Lindegren2016} reported a relatively small offset of
$-0.089$ mas between DR1 and the revised {\it Hipparcos} results, they
found a north-south (N-S) asymmetry when dividing their sample at
ecliptic latitude ($\beta$) = 0: the northern stars have a larger
offset ($-0.13$ mas) than the southern stars ($-0.05$ mas).  We
examined our sample of 612 stars within 25 pc for a N-S asymmetry, and
also find a significant difference, {\it but in the opposite sense of
  \cite{Lindegren2016}}: the southern star parallaxes are offset by
$-0.32\pm0.04$, whereas the northern stars are offset by only
$-0.17\pm0.03$ (see Table~\ref{tbl:NS}).  There is no obvious
explanation for this difference.

% We found that the larger offset in the south may be skewed by the
% number of stars within 80 mas in the south, not by their locations
% on the sky.

%%%%%%%%%%%%%%%%%%%%%%%%%%%%%%%%%%%%%%%%%%%%%%%%%%%%%%%%%%%%%%%%%%%%
\section{Conclusions}
%%%%%%%%%%%%%%%%%%%%%%%%%%%%%%%%%%%%%%%%%%%%%%%%%%%%%%%%%%%%%%%%%%%%

We have performed parallax comparisons between the {\it Gaia's} DR1
and previously available parallax measurements for single stars within
25 pc.  We conclude that there is an offset of $-0.24\pm0.02$ mas
between DR1 and available weighted mean parallaxes, in the sense that
the DR1 results place stars at slightly larger distances than previous
measurements. We also found that the Luz-Kelker bias has a negligible
effect on this systematic offset reported here. Although the offset is
small, it is particularly important to resolve this discrepancy if it
systematically affects stars at much greater distances, for which the
parallax measurements will be less than a few milliarcseconds.  We
also found that the offset depends on how the samples are drawn.
Nearby stars tend to have larger offsets than more distant stars, at
least out to 100 pc.  The offsets for the ensemble of stars within 25
pc is comparable to that reported in \cite{Stassun2016}, yet only 10\%
of their binaries were closer than 100 pc.  We suspect that they have
over-estimated their offset because most of their stars are much
further away than our 612 single stars.  Following the trend of
smaller offsets at larger distances, more distant samples, like that
in \cite{Lindegren2016}, will exhibit smaller offsets. However, we
expect the next {\it GAIA} data release will resolve this offset due
to distances. Given the significant work that has gone into measuring
their large parallaxes independently, the nearest stars, once again,
can play a crucial role in understanding our Milky Way.

%%%%%%%%%%%%%%%%%%%%%%%%%%%%%%%%%%%%%%%%%%%%%%%%%%%%%%%%%%%%%%%%%%%%
\section{Acknowledgments}
%%%%%%%%%%%%%%%%%%%%%%%%%%%%%%%%%%%%%%%%%%%%%%%%%%%%%%%%%%%%%%%%%%%%
We thank the anonymous referee and N. Hambly for valuable comments
that improved this manuscript. Work on constructing this 25 pc sample
has been supported by National Science Foundation under grants
AST-0507711, AST-0908402, AST-1109445, and AST-141206.  This research
has made use of the SIMBAD database, operated at CDS, Strasbourg,
France.  This publication makes use of data products from the Two
Micron All Sky Survey, which is a joint project of the University of
Massachusetts and the Infrared Processing and Analysis
Center/California Institute of Technology, funded by the National
Aeronautics and Space Administration and the National Science
Foundation.  This work has made use of data from the European Space
Agency (ESA) mission {\it Gaia}
(\url{http://www.cosmos.esa.int/gaia}), processed by the {\it Gaia}
Data Processing and Analysis Consortium
(DPAC,\\ \url{http://www.cosmos.esa.int/web/gaia/dpac/consortium}). Funding
for the DPAC has been provided by national institutions, in particular
the institutions participating in the {\it Gaia} Multilateral
Agreement.

%%%%%%%%%%%%%%%%%%%%%%%%%%%%%%%%%%%%%%%%%%%%%%%%%%%%%%%%%%%%%%%%%%%%

%%%%%%%%%%%%%%%%%%%%%%%%%%%%%%%%%%%%%%%%%%%%%%%%%%%%%%%%%%%%%%%%%%%%

%\clearpage

%%%%%%%%%%%%%%%%%%%%%%%%%%%%%%%%%%%%%%%%%%%%%%%%%%%%%%%%%%%%%%%%%%%%
%                             Table
%%%%%%%%%%%%%%%%%%%%%%%%%%%%%%%%%%%%%%%%%%%%%%%%%%%%%%%%%%%%%%%%%%%%

\begin{deluxetable}{cclllrrcrcrrrrccrr}
\setlength{\tabcolsep}{0.02in}
\rotate
\tablewidth{0pt}
\tabletypesize{\tiny}
\tablehead{\colhead{R.A. (2000.0)}             &
           \colhead{Decl. (2000.0)}            &
           \colhead{Name}             &
           \colhead{HIP}              &
           \colhead{Gaia}             & % source_id
           \colhead{$\mu$}            &
           \colhead{P.A.}             &
           \colhead{Ref}              &
           \colhead{V}                &
           \colhead{Ref}              &
           \colhead{$K_{s}$}          &
           \colhead{G}                &
           \colhead{$\varpi_{weighted}$}  &
           \colhead{$\varpi_{err}$}       &
           \colhead{\# of $\varpi$}      &
           \colhead{Ref}              &
           \colhead{$\varpi_{Gaia}$}      &
           \colhead{$\varpi_{err}$}      \\
           \colhead{}                 &
           \colhead{}                 &
           \colhead{}                 &
           \colhead{}                 &
           \colhead{source\_id}       &
           \colhead{(\arcsec/yr)}     &
           \colhead{(deg)}            &
           \colhead{}                 &
           \colhead{(mag)}            &
           \colhead{}                 &
           \colhead{(mag)}            &
           \colhead{(mag)}            &
           \colhead{(mas)}            &
           \colhead{(mas)}            &
           \colhead{}                 &
           \colhead{}                 &
           \colhead{(mas)}            &
           \colhead{(mas)}            \\
           \colhead{(1)}              &
           \colhead{(2)}              &
           \colhead{(3)}              &
           \colhead{(4)}              &
           \colhead{(5)}              &
           \colhead{(6)}              &
           \colhead{(7)}              &
           \colhead{(8)}              &
           \colhead{(9)}              &
           \colhead{(10)}             &
           \colhead{(11)}             &
           \colhead{(12)}             &
           \colhead{(13)}             &
           \colhead{(14)}             &
           \colhead{(15)}             &
           \colhead{(16)}             &
           \colhead{(17)}             &
           \colhead{(18)}             }
\startdata
00:05:17.68  & $-$67:49:57.4 & GJ0003          &436        &4706630496753986944     & 0.571 & 191.9 & 8 &  8.48 & 2  & 5.71 &  8.00 &  62.970 & 0.709 & 2 & 23,26    &  61.142 & 0.3358 \\         
00:06:19.19  & $-$65:50:25.9 & LHS1019         &523        &4899957901143352576     & 0.586 & 160.5 & 8 & 12.17 & 13 & 7.63 & 10.92 &  59.850 & 2.640 & 1 & 26       &  57.100 & 0.3270 \\               
00:06:36.77  & $+$29:01:17.4 & GJ0005          &544        &2860924685628241024     & 0.420 & 115.1 & 8 &  6.06 & 1  & 4.31 &  5.82 &  73.100 & 0.558 & 2 & 23,26    &  72.627 & 0.5204 \\         
00:08:27.29  & $+$17:25:27.3 & HIP000687       &687        &2772804841615919616     & 0.110 & 233.8 & 8 & 10.80 & 19 & 6.98 &  9.89 &  45.980 & 1.930 & 1 & 26       &  46.386 & 0.3303 \\               
00:12:50.25  & $-$57:54:45.4 & GJ2001          &1031       &4918476357015844352     & 0.123 & 286.8 & 8 &  7.23 & 13 & 5.38 &  6.96 &  49.530 & 0.580 & 1 & 26       &  48.461 & 0.3542 \\               
\enddata
\tablerefs{ 
(1)this work;
(2)\citealt{Bes90};
(3)\citealt{Dah88};
(4)\citealt{Dit14};
(5)\citealt{Fab00};
(6)\citealt{Har80};
(7)\citealt{Har93};
(8)\citealt{Hog2000};
(9)\citealt{Ian96};
(10)\citealt{Jao2005};
(11)\citealt{Kho13};
(12)\citealt{Kil07};
(13)\citealt{Koe10};
(14)\citealt{Rie10};
(15)\citealt{Shkolnik2012};
(16)\citealt{Subasavage2009};
(17)\citealt{Wei87};
(18)\citealt{Wei88};
(19)\citealt{Wei93};
(20)\citealt{Wei96}
(21)\citealt{Wei99};
(22)\citealt{Win15};
(23)\citealt{YPC};
(24)\citealt{Wei91a};
(25)\citealt{Wei91b};
(26)\citealt{Leeuwen2007}
}
\tablecomments{This table is available in its entirety in a machine-readable form in the online journal. A portion is shown here for guidance regarding its form and content.}
\label{tbl:data}
\end{deluxetable}

\begin{deluxetable}{rcccccc}
\setlength{\tabcolsep}{0.06in}
\rotate
\tablewidth{0pt}
\tabletypesize{\small}
\tablehead{
           \colhead{}              &
           \multicolumn{2}{c}{Method 1} &
           \multicolumn{2}{c}{Method 2} &
           \multicolumn{2}{c}{Method 3} \\
           \colhead{}              &
           \multicolumn{2}{c}{weighted mean horizon} &
           \multicolumn{2}{c}{{\it Gaia}-DR1 horizon} &
           \multicolumn{2}{c}{{\it Gaia}-DR1 horizon} \\
           \hline
           \colhead{}              &
           \multicolumn{2}{c}{$\varpi_{GAIA-weighted}$} &
           \multicolumn{2}{c}{$\varpi_{GAIA-weighted}$} &
           \multicolumn{2}{c}{$\varpi_{GAIA-vLe07}$} \\
           \hline
           \colhead{parallax sets} &
           \colhead{\# of stars}   &
           \colhead{offset}        &
           \colhead{\# of stars}   &
           \colhead{offset}        &
           \colhead{\# of stars}   &
           \colhead{offset}        \\
           \colhead{(mas)}         &
           \colhead{}              &
           \colhead{(mas)}         &
           \colhead{}              &
           \colhead{(mas)}         &
           \colhead{}              &
           \colhead{(mas)}         }

\startdata
   $\varpi>$80    &     85 &  $-$0.34$\pm$0.11 &     80 & $-$0.29$\pm$0.10  &    128 &   $-$0.50$\pm$0.08  \\
40$<\varpi\leq$80 &    527 &  $-$0.23$\pm$0.02 &    500 & $-$0.17$\pm$0.02  &    725 &   $-$0.16$\pm$0.03  \\
20$<\varpi\leq$40 & \nodata&    \nodata        &\nodata &   \nodata         &   3815 &   $-$0.08$\pm$0.02  \\
10$<\varpi\leq$20 & \nodata&    \nodata        &\nodata &   \nodata         &  11592 &   $-$0.06$\pm$0.01  \\
\hline                                                               
   all            &    612 &  $-$0.24$\pm$0.02 &    580 & $-$0.19$\pm$0.02  &  16260 &   $-$0.07$\pm$0.01  \\
\enddata
\label{tbl:offset}
\end{deluxetable}

%%%%% North-South sample

\begin{deluxetable}{ccccc}
\setlength{\tabcolsep}{0.06in}
\tablewidth{0pt}
\tabletypesize{\small}
\tablehead{
     \colhead{}        &
     \colhead{offset}  &
     \colhead{$\varpi_{weighted}>$ 80 mas} &
     \colhead{80 $\geq\varpi_{weighted}>$ 40 mas} &
     \colhead{total}\\
     \colhead{}        &
     \colhead{(mas)}  &
     \colhead{\# of stars} &
     \colhead{\# of stars} &
     \colhead{\# of stars}
      }
\startdata
North ($\beta>$ 0.)    &  $-$0.17$\pm$0.03  &  39 & 266  & 305 \\
South ($\beta<$ 0.)    &  $-$0.32$\pm$0.04  &  46 & 261  & 307 \\
\enddata
\label{tbl:NS}
\end{deluxetable}

\end{document}